\documentclass[aps,prd,showpacs,amssymb,nofootinbib]{revtex4-1}
\usepackage{amsmath,amsthm,amsfonts}

\newcommand{\be}{\begin{equation}}
\newcommand{\ee}{\end{equation}}
\usepackage{hyperref}

\begin{document}

\title{Further Improvements in the Understanding of Isotropic Loop
Quantum Cosmology}

\author{M. Mart\'{i}n-Benito}
\email{merce.martin@iem.cfmac.csic.es} \affiliation{Instituto de
Estructura de la Materia, CSIC, Serrano 121, 28006 Madrid, Spain}
\author{G. A. Mena Marug\'{a}n}\email{mena@iem.cfmac.csic.es}
\affiliation{Instituto de Estructura de la Materia, CSIC, Serrano
121, 28006 Madrid, Spain}
\author{J. Olmedo}\email{olmedo@iem.cfmac.csic.es}
\affiliation{Instituto de Estructura de la Materia, CSIC, Serrano
121, 28006 Madrid, Spain}

\begin{abstract}

The flat, homogeneous, and isotropic universe with a
massless scalar field is a paradigmatic model in Loop
Quantum Cosmology. In spite of the prominent role that
the model has played in the development of this branch
of physics, there still remain some aspects of its
quantization which deserve a more detailed discussion.
These aspects include the kinematical resolution of
the cosmological singularity, the precise relation
between the solutions of the densitized and
non-densitized versions of the quantum Hamiltonian
constraint, the possibility of identifying
superselection sectors which are as simple as
possible, and a clear comprehension of the
Wheeler-DeWitt (WDW) limit associated with the theory
in those sectors. We propose an alternative operator
to represent the Hamiltonian constraint which is
specially suitable to deal with all these issues in a
detailed and satisfactory way. In particular, with our
constraint operator, the singularity decouples in the
kinematical Hilbert space and can be removed already
at this level. Thanks to this fact, we can densitize
the quantum Hamiltonian constraint in a
well-controlled manner. Besides, together with the
physical observables, this constraint superselects
simple sectors for the universe volume, with a
discrete support contained in a single semiaxis of the
real line and for which the basic functions that
encode the information about the geometry possess
optimal physical properties. Namely, they provide a
no-boundary description around the cosmological
singularity and admit a well-defined WDW limit in
terms of standing waves. Both properties explain the
presence of a generic quantum bounce replacing the
classical singularity at a fundamental level, in
contrast with previous studies where the bounce was
proved in concrete regimes --focusing on states with a
marked semiclassical behavior-- or for a simplified
model.

\end{abstract}


\maketitle

\section{Introduction}
\label{sec:intro}

Loop Quantum Gravity (LQG) \cite{lqg1,lqg2,lqg3} is
one of the most promising candidates for a theory of
quantum gravity. Nonetheless, although this
quantization program of General Relativity has been
extensively developed, it has not been completed yet.
In order to test the program and obtain physical
predictions in simple situations of interest, Loop
Quantum Cosmology (LQC) implements the quantization
procedures of LQG in symmetry reduced models
\cite{lqc}. This field has suffered a major
breakthrough in recent years thanks to the large
number of studies carried out in models such as flat
FRW cosmologies \cite{abl,aps1,aps3,acs,cs,kale, Pol},
FRW universes with other topologies
\cite{apsv,vand,skl} or with a non-vanishing
cosmological constant \cite{tom}, anisotropic Bianchi
I models \cite{chio,mmp,mmp2,awe}, or even some
inhomogeneous scenarios \cite{gow}.

In this work, we will focus our attention on the flat
FRW model with a massless scalar field, which is the
most prominent system studied so far in LQC. In fact,
even though this is the simplest cosmological model
with non-trivial dynamics, its polymeric quantization
has already provided relevant results
\cite{abl,aps1,aps3}, such as the validity of the
classical dynamics for semiclassical states in LQC in
the region far away from the classical singularity
and, more importantly, the replacement of this big
bang singularity by a quantum bounce. In addition, the
robustness of these results has been carefully checked
\cite{acs,cs}. Besides, those studies, together with
other complementary analyses \cite{kale,Pol}, have
served to rigorously establish the mathematical
foundations of LQC.

Nevertheless, despite the extensive analysis of this
model performed up to date, there are still some
issues which are not completely clear and need a more
careful discussion in order to have a thorough
understanding of the theory. In particular, the points
that we have in mind involve, on the one hand, a
rigorous densitization of the Hamiltonian constraint
(with respect to the volume of the universe) at the
quantum level --a step which is carried out in order
to obtain a \emph{densitized} Hamiltonian constraint
that is easier to solve than the \emph{non-densitized}
one arising naturally from LQG--, and, on the other
hand, the superselection of the kinematical Hilbert
space in different sectors which are intended to be as
simple as possible while possessing optimal physical
properties. Among such properties, we are interested
in the existence of a regime with a well-defined WDW
limit and where general features of the states provide
a basis to explain the occurrence of a quantum bounce.

With the aim of investigating these questions while
keeping a rigorous mathematical control of all the
steps in the discussion, we will adopt a new
prescription when promoting the Hamiltonian constraint
of the cosmological model to a symmetric operator.
Actually, our quantum Hamiltonian constraint results
more suitable than previous proposals inasmuch as it
clarifies the issues stated above. One of the
noticeable features of our prescription is the
resolution of the cosmological singularity already at
the kinematical level, in the sense that our
Hamiltonian constraint allows us to decouple the
zero-volume state from the rest of states in the
kinematical Hilbert space, so that the kernel of the
volume operator can be removed from the quantum theory
(in a certain sense, this implements ideas from
Bojowald \cite{boj}). Once the singularity has been
eliminated, we are able to \textit{formally} establish
a bijection between the solutions of the densitized
and non-densitized versions of the quantum Hamiltonian
constraint, an aspect which was not properly
considered in previous analyses. Another important
feature of our Hamiltonian constraint is that it
decouples states with opposite orientations of the
triad. Owing to this fact, we can identify in our
theory superselection sectors which are simpler than
those obtained in Refs. \cite{aps3,acs}, since they
are supported in discrete semilattices contained in a
single semiaxis of the real line, instead of the whole
line. Therefore, in our theory we can restrict the
study to (e.g.) the positive semiaxis in a natural
way, without the need to introduce any particular
boundary condition or appeal to the parity symmetry
discussed in Ref. \cite{aps3} (this symmetry is
straightforward to impose in our case). The simplicity
of our superselection sectors allows us to gain
physical intuition and improve our knowledge about the
physical consequences of the loop quantization. On the
one hand, the basic elements which encode the
information about the geometry, namely the generalized
eigenfunctions of the gravitational part of the
constraint (defined as a self-adjoint operator), are
each explicitly determined by a single piece of
initial data. They ``arise'' in a single section of
minimum non-zero volume, without crossing the
singularity to the sector of opposite triad
orientation and without satisfying any kind of
boundary condition. In this sense, the
self-adjointness of the gravitational part of the
constraint leads to a no-boundary description. On the
other hand, whereas this part of the constraint is an
operator with non-degenerate spectrum, its analog in
the WDW theory is two-fold degenerate. This property,
together with the fact that it is a real difference
operator in the volume representation, implies that
its eigenfunctions have a WDW limit with the form of
an exact standing wave, i.e., each eigenfunction
converges to a specific linear combination of the two
analog WDW elements (one of them contracting in volume
and the other one expanding) where they both
contribute with equal amplitudes. Since, owing to the
no-boundary property, the outgoing flux associated
with the expanding component must become incoming flux
corresponding to the contracting component, and vice
versa, a quantum bounce must happen. In this way, we
will be able to explain the existence of a quantum
bounce, showing that it is a direct consequence of the
quantum geometry underlying LQC and hence a
fundamental feature of the theory.

The paper is organized as follows. In Sec.
\ref{sec:kin} we review the classical framework and
the quantum kinematics. We choose a new, suitable
symmetric ordering for the Hamiltonian constraint and
densitize it in Sec. \ref{sec:ham}. The gravitational
part of the constraint is analyzed in detail in Sec.
\ref{sec:grav}. We then impose the constraint in Sec.
\ref{sec:phys}, obtaining the physical Hilbert space.
In Sec. \ref{sec:phys2} we compare different aspects
of our proposal and of previous ones, showing the
advantages of our approach. Finally, in Sec.
\ref{sec:dis} we discuss the main conclusions of our
analysis.

\section{The Flat, Homogeneous, and Isotropic Model}
\label{sec:kin}

We briefly summarize here the classical formulation of
the model and its kinematical treatment in LQC. For
more details, see Refs. \cite{abl,aps3}.

\subsection{The Classical Approach}
\label{subsec:classic}

In the flat model under study, the spatial
hypersurfaces are non-compact. Therefore, in
principle, the spatial integrals arising in this
system diverge. In order to avoid this problem, one
introduces a Euclidean fiducial 3-metric and restrict
the integrals to a given cell, which is cubic with
respect to the fiducial metric and whose fiducial
volume is $V_0$ \cite{abl}. Owing to the imposition of
homogeneity, the spatial diffeomorphism freedom is
fixed and the associated constraints are trivially
satisfied. Furthermore, the internal [SU(2)] gauge
freedom is removed by choosing a diagonal gauge, in
which the connection and the densitized triad are
respectively given by \be
A^i_a=cV_0^{-1/3}\delta_a^i\,\,\,\textrm{and}\,\,\,
E^a_i=pV_0^{-2/3}\,\delta^a_i.\ee So, one can identify
the internal indices $i=1,2,3$ with those for the
tangent space $a=1,2,3$, and the three directions are
equivalent owing to the isotropy. In this way, the two
quantities $c$ and $p$ represent the only degrees of
freedom of the geometry. This choice of
parameterization for $A^i_a$ and $E^a_i$ leaves all
the relevant physical quantities independent of the
fiducial structures, as it is the case for the
symplectic structure, determined by the Poisson
brackets $\{c,p\}=8\pi G\gamma/3$, where $G$ is the
Newton constant and $\gamma>0$ is the Immirzi
parameter.

In order to have non-trivial dynamics, it is necessary
to add matter degrees of freedom. The simplest
possibility is a homogeneous massless scalar field
$\phi$, with conjugate momentum $p_\phi$, such that
$\{\phi,p_\phi\}=1$.

The model is then subject only to the Hamiltonian
constraint. Making use of the spatial homogeneity, the
integrated form of this constraint for any lapse
function $N$ is $\mathbb{C}(N)=NC$, where the
Hamiltonian constraint \be
\label{class-const}C=-\frac{6}{\gamma^{2}}c^2\sqrt{|p|}
+8\pi G\frac{p^2_\phi}{|p|^{3/2}}\ee has the standard
densitization used in LQG. It is easy to check that,
classically, both $p$ and $\phi$ are monotonous
functions of the proper time. Usually, $\phi$ is
regarded as an internal time, with all quantities
evolving as functions of it. Then, the classical
solutions represent expanding or contracting
universes, and all of them are singular.

\subsection{Quantum Representation}
\label{subsec:qrepres}

In LQG, the gauge invariant information about the
phase space is captured in holonomies of
su(2)-connections and fluxes through surfaces.
Similarly, in LQC one adopts as basic variables
holonomies along straight edges of oriented coordinate
length $\mu V_0^{1/3}$ in the fiducial directions
--that are given by $h_i^{\mu}(c)=e^{\mu c\tau_i}$ for the
direction $i$ \cite{note1}-- and fluxes through
squares normal to these fiducial directions --which
are equal to $p$ up to unimportant factors. The
configuration algebra $\textrm{Cyl}_S$ is the algebra
of almost periodic functions of $c$ \cite{Vel, Castro}
generated by the matrix elements of the holonomies,
$\mathcal{N}_\mu=e^{i\mu c/2}$. In the momentum
representation, the states defined by these matrix
elements are denoted by $|\mu\rangle$. The Cauchy
completion of $\textrm{Cyl}_S$ with respect to the
discrete norm
$\langle\mu|\mu'\rangle=\delta_{\mu\mu'}$ provides the
gravitational part of the kinematical Hilbert space
$\mathcal{H}^{\textrm{grav}}_{\textrm{kin}}$. The
operator $\hat{p}$ has a diagonal action on the basis
formed by the states $|\mu\rangle$, whereas
$\hat{\mathcal{N}}_\mu$ shifts the state
$|\mu'\rangle$ to $|\mu'+\mu\rangle$.

The standard procedure in LQC to define the curvature
of the su(2)-connection is to express it in terms of
holonomies along closed loops and shrink their area to
the minimum non-zero value $\Delta$ allowed by LQG,
which is proportional to the Immirzi parameter and to
the square of the Planck length
$l_{\textrm{Pl}}=\sqrt{G\hbar}$. Then, the condition
that the physical area of a square with sides of
(minimum) fiducial length $\bar\mu V_0^{1/3}$ equals
$\Delta$ \cite{aps3} leads to the operator relation
\be\widehat{\frac{1}{\bar{\mu}}}=
\frac{\widehat{\sqrt{|p|}}}{\sqrt{\Delta}}.\ee Since
the corresponding shift produced by
$\hat{\mathcal{N}}_{\bar{\mu}}$ is not constant in the
basis of states $|\mu\rangle$, it is convenient to
relabel these states by introducing an affine
parameter $v(\mu)$, which is proportional to the
respective eigenvalue of the physical volume operator.
In this manner, $\hat{\mathcal{N}}_{\bar\mu}$ produces
a constant increment in the volume,
$\hat{\mathcal{N}}_{\bar{\mu}}|v\rangle=|v+1\rangle$,
while $\hat{p}|v\rangle=\textrm{sign}(v)(2 \pi \gamma
l_{\textrm{Pl}}^2\sqrt{\Delta}| v|)^{2/3}|v\rangle$.

On the other hand, for the matter part of the
kinematical Hilbert space one takes the standard
representation space of square integrable functions
with respect to the Lebesgue measure, namely
$\mathcal{H}^{\textrm{mat}}_{\textrm{kin}}=
L^2(\mathbb{R},d\phi)$. In total, the kinematical
Hilbert space is $\mathcal{H}_{\textrm{kin}}=
\mathcal{H}^{\textrm{grav}}_{\textrm{kin}}
\otimes\mathcal{H}^{\textrm{mat}}_{\textrm{kin}}$.

\section{The Hamiltonian Constraint Operator}
\label{sec:ham}

Once the kinematical structure of the quantization has
been introduced, we can proceed to represent the
Hamiltonian constraint $C$ as an operator. We will
choose a symmetric ordering for the constraint which
differs from the orderings considered in Refs.
\cite{aps3,acs} but, as we will see, results more
suitable.

As usual in LQC, to represent the factor $|p|^{-3/2}$
in the second term of Eq. \eqref{class-const} one
appeals to Thiemann's procedure \cite{lqg3}, rewriting
it in terms of Poisson brackets of holonomies with the
physical volume $V:=|p|^{3/2}$. On the other hand, in
the gravitational part of the constraint, one
expresses the curvature of the su(2)-connection using
holonomies as explained above, in order to obtain a
well defined operator \cite{abl,aps1,aps3}.

In earlier works, the quantum Hamiltonian constraint
was constructed along these lines, adopting certain
choices of factor ordering \emph{directly} for the
isotropic model \cite{abl,aps1,aps3,acs}. In contrast,
in our case, we will adopt an ordering motivated from
previous studies performed in the anisotropic Bianchi
I model \cite{mmp}. In such a system, the sign of the
triad plays an important role because there are three
different directions, rather than one. Then, products
of two signs are not necessarily equal to the unity.
As a consequence, in that case, one has to take
carefully into account the presence of the signs when
symmetrizing the constraint. To select a similar
ordering in the present isotropic case, we start with
the symmetric Hamiltonian constraint constructed in
Ref. \cite{mmp} and identify the three spatial
directions. In this manner we arrive to the following
constraint operator \be
\label{grav-cons}\hat{C}:=\left[\widehat{
\frac{1}{V}}\right]^{1/2}\left(-\frac{6}{\gamma^{2}}
\widehat{\Omega}^2+8\pi
G\hat{p}_{\phi}^2\right)\left[\widehat{\frac{1}{V}}
\right]^{1/2}.\ee Here, the inverse volume operator is
given by
\begin{align}\label{inv-volum}
 \left[\widehat{\frac{1}{V}}\right]&:=
\left[\widehat{\frac{1}{\sqrt{|p|}}}\right]^{3},
\end{align}
where
\begin{align}
\widehat{\frac{1}{\sqrt{|p|}}}&=\frac{3}{4\pi\gamma
l_{\textrm{Pl}}^2\sqrt{\Delta}}
\widehat{\text{sign}(p)}
\widehat{\sqrt{|p|}}\left(\hat{\mathcal
N}_{-\bar{\mu}}\widehat{\sqrt{|p|}}\hat{\mathcal
N}_{\bar{\mu}}- \hat{\mathcal
N}_{\bar{\mu}}\widehat{\sqrt{|p|}}\hat{\mathcal
N}_{-\bar{\mu}}\right).
\end{align}
The inverse volume acts
diagonally on the basis states $|v\rangle$ and
annihilates the state $|v=0\rangle$.

On the other hand the operator
\begin{align}\label{grav-op}
\widehat\Omega&:=\frac1{4i\sqrt{\Delta}}
\left[\widehat{\frac1{\sqrt{|p|}}}\right]^{-1/2}
\widehat{\sqrt{|p|}} \left[\left(\hat{\mathcal
N}_{2\bar\mu}-\hat{\mathcal
N}_{-2\bar\mu}\right)\widehat{\text{sign}(p)}
+\widehat{\text{sign}(p)}\left(\hat{\mathcal
N}_{2\bar\mu}-\hat{\mathcal
N}_{-2\bar\mu}\right)\right]
\widehat{\sqrt{|p|}}\left[\widehat{
\frac1{\sqrt{|p|}}}\right]^{-1/2}
\end{align}
coincides (in terms of holonomies and triad
operators) with the one studied in Ref. \cite{mmp},
denoted there by $\hat{\Theta}_i$. The factors
involving powers of the triad at the beginning and the
end of this expression appear in a very particular
manner, in analogy to the Bianchi I model \cite{mmp}.
Nonetheless, this concrete ordering is not relevant
inasmuch as physical results do not depend appreciably
on it, and one could select any other one. For
instance, one could even ignore the quantum effects
coming from the inverse of the volume \cite{notepl}, as
was done in Ref. \cite{acs}. The really relevant
features of our ordering, as we will see in the next
section, come from the particular treatment of the
sign of $p$ that we have considered.

It is straightforward to see that our Hamiltonian
constraint operator \textit{annihilates} the state
$|v=0\rangle$ and leaves invariant its orthogonal
complement, which we will denote by
$\widetilde{\mathcal{H}}_{\textrm{kin}}^{\textrm{grav}}$.
Hence, when studying the non-trivial solutions of the
Hamiltonian constraint, we can restrict ourselves
to this latter subspace. Note that
$\widetilde{\mathcal{H}}_{\textrm{kin}}^{\textrm{grav}}$
is just the Cauchy completion with the discrete norm
of $\widetilde{\textrm{Cyl}}_S$, the linear span of
all the $|v\rangle$ states with $v\neq0$. As in
similar situations studied by us in previous works
\cite{mmp,mmp2,gow,Castro}, the big bang is then
resolved in the sense that the quantum equivalent to
the classical singularity (namely, the eigenstate of
vanishing physical volume) has been entirely removed
from the kinematical Hilbert space.

Once we have removed the kernel of the inverse volume
operator, we can introduce the densitized version of
$\hat C$. Let us remember before that the physical
states, which are annihilated by the Hamiltonian
constraint, are generally not normalizable in the
kinematical Hilbert space. We should seek them in a
larger space. A natural home, as far as the
gravitational part of the system is concerned, is the
algebraic dual $\widetilde{\textrm{Cyl}}_S^*$ of the
dense set $\widetilde{\textrm{Cyl}}_S$. In this dual
space (tensor product any suitable space for the
matter degrees of freedom), we can establish a
one-to-one relation between any element $(\psi|$
annihilated by the (adjoint of the) operator $\hat{C}$
and any other element
$(\psi'|=(\psi|[\widehat{1/V}]^{1/2}$ annihilated by
the (adjoint of the) densitized version of the
constraint, which is given by
\be\label{den-const}\hat{\mathcal{C}}=-\frac{6}{\gamma^{2}}
\widehat{\Omega}^2+8\pi G\hat{p}_{\phi}^2.\ee This
equivalent form of the Hamiltonian constraint is
easier to impose since obviously $\widehat{\Omega}^2$
and $\hat{p}_{\phi}^2$ are Dirac observables which
commute.

Another way to get an easily solvable Hamiltonian
constraint is to directly promote the classical
densitized constraint $\mathcal C:=VC$ to an operator.
Nevertheless, as we have commented, such an object
does not arise from the standard densitization of the
constraint in LQG. For this reason, it seems natural
to respect this latter densitization and show that one
can construct a bijection between the solutions to the
two considered Hamiltonian constraints. We emphasize
that this bijection cannot be established in the
kinematical Hilbert space, both because the physical
volume operator is unbounded, and therefore it is not
defined in the whole space, and because solutions do
not belong to this space indeed.

\section{Characterization of the Gravitational
Constraint Operator}\label{sec:grav}

As we have already seen, the operator
$\widehat\Omega^2$, which provides the gravitational
part of the Hamiltonian constraint \eqref{den-const},
is a Dirac observable. Since the matter part of this
Hamiltonian constraint is well known, the study of the
properties of $\widehat\Omega^2$, and in particular
its spectral analysis, is the essential step for the
resolution of the constraint. In this section, we will
carry out a detailed analysis of this operator. We
will identify first the superselection sectors which
arise in the theory. Afterwards, we will perform the
spectral analysis of $\widehat\Omega^2$ in those
sectors and determine its eigenfunctions. Remarkably,
we can obtain their explicit expression, what will
allow us to study their behavior. On the one hand, we
will discuss how these eigenfunctions realize the
commented no-boundary description. On the other hand,
we will relate them with the eigenfunctions of the
geometrodynamical (WDW) counterpart of
$\widehat\Omega^2$ for large $v$, in order to
understand the WDW limit of our theory. Both features
will give us insights about the existence of a quantum
bounce. As a complement, we will also relate
$\widehat\Omega^2$ with the operator $\widehat\Omega$.
Finally, we will compare our results with those of
previous works on the polymeric quantization of the
model, to point out the goodness of our prescription.

\subsection{Superselection Sectors}
\label{subsec:superselect}

In LQC, owing to the discreteness of the volume
representation, the gravitational part of the
Hamiltonian constraint turns out to be a difference
operator instead of a differential one, as happens to
be the case in the standard WDW theory. In the
particular case of isotropic cosmologies, this
operator produces a shift of four units in the label
of the basis states $|v\rangle$. As a consequence,
only basis states with support in discrete lattices of
step four are related under its action. In addition,
the Hilbert spaces of states which have support in
such lattices turn out to be superselected, inasmuch
as they are also preserved under the action of the
complete Hamiltonian constraint and the physical
observables. Whereas these results are general within
isotropic LQC, in this subsection we will see that the
lattices associated with our operator
$\widehat\Omega^2$, and the corresponding
superselection sectors, are simpler than those
obtained in previous studies of the model.

The action of $\widehat\Omega^2$ on the basis states
$|v\rangle$ of
$\widetilde{\mathcal{H}}^{\textrm{grav}}_{\textrm{kin}}$
takes the form
\begin{align}\label{grav-op-action}
\widehat\Omega^2|v\rangle&=-f_+(v)f_+(v+2)|v+4\rangle+
\left[f_+^2(v)+f_-^2(v)\right]|v\rangle-f_-(v)f_-(v-2)
|v-4\rangle,
\end{align}
where
\begin{eqnarray}\label{f}
f_\pm(v)&:=&\frac{\pi\gamma\l_{\textrm{Pl}}^2}{3}
g(v\pm2)s_\pm(v)g(v), \\\label{s}
s_\pm(v)&:=&\text{sign}(v\pm2)+\text{sign}(v),
\end{eqnarray}
and
\begin{align}\label{g}
g(v)&:=
\begin{cases}
\left|\left|1+\frac1{v}\right|^{\frac1{3}}
-\left|1-\frac1{v}\right| ^{\frac1{3}}
\right|^{-\frac1{2}} & {\text{if}} \quad v\neq 0,
\vspace*{.2cm}\\
0 & {\text{if}} \quad v=0.\\
\end{cases}
\end{align}

Notice that the combination of signs in the functions
$f_\pm(v)$, which comes from the factor ordering that
we have chosen to symmetrize the constraint operator,
is the responsible of a remarkable property, namely,
$f_-(v)f_-(v-2)=0$ if $v\in(0,4]$, while
$f_+(v)f_+(v+2)=0$ for $v\in[-4,0)$. Therefore, the
positive and negative semiaxes, $v>0$ and $v<0$, are
decoupled under the action of our operator
$\widehat\Omega^2$, as one can see from Eq.
\eqref{grav-op-action}. In conclusion, this operator
relates only basis states $|v\rangle$ with $v$
belonging to one of the semilattices of step four
$\mathcal
L_{\tilde\varepsilon}^\pm:=\{v=\pm(\tilde\varepsilon+4n),
\,n\in\mathbb{N}\}$, where
$\tilde\varepsilon\in(0,4]$. In other words,
$\widehat\Omega^2$ is well defined in any of the
Hilbert spaces $\mathcal H^\pm_{\tilde\varepsilon}$
obtained as the closure of the respective domains
$\text{Cyl}_{\tilde\varepsilon}^{\pm}:=\text{span}
\{|v\rangle,\,v\in\mathcal
L_{\tilde\varepsilon}^\pm\}$ with respect to the discrete
inner product. Note that the
non-separable kinematical Hilbert space $
\widetilde{\mathcal H}^{\text{grav}}_{\text{kin}}$ can
be written as a direct sum of separable subspaces in
the form $\widetilde{\mathcal
H}^{\text{grav}}_{\text{kin}}=\oplus_{\tilde\varepsilon}
(\mathcal H^+_{\tilde\varepsilon}\oplus\mathcal
H^-_{\tilde\varepsilon})$.

The Hilbert spaces $\mathcal H^\pm_{\tilde\varepsilon}
\otimes L^2(\mathbb{R},d\phi)$ are preserved by the
action of the Hamiltonian constraint (and then, as we
will see, of physical observables). Thus, these
Hilbert spaces provide superselection sectors in our
theory. We can restrict the study to any of them. For
concreteness, in the following we will work in
$\mathcal H^+_{\tilde\varepsilon}\otimes
L^2(\mathbb{R},d\phi)$.

The difference with respect to previous works on the
polymeric quantization of this model
\cite{abl,aps1,aps3,acs,kale,Pol} is that the support
of our sectors is contained in a single semiaxis of
the real line, while in those works the sectors had
contributions both from the positive and the negative
semiaxes, which were not decoupled. Later on, we will
discuss the advantages of working with our simpler
sectors.

\subsection{Spectral Analysis}
\label{subsec:spectral}

Let us begin by showing that the symmetric operator
$\widehat\Omega^2$ (with domain
$\text{Cyl}_{\tilde\varepsilon}^{+}$) is essentially
self-adjoint. Actually, one can calculate that the
difference between $\alpha \widehat\Omega^2$, with
$\alpha:=3/(4\pi\gamma^2 l_\text{Pl}^2\hbar),$ and the
gravitational constraint operator $H'_\text{APS}$ of
Ref. \cite{kale}, both being defined in the Hilbert
space $\mathcal H^+_{\tilde\varepsilon}\oplus\mathcal
H^-_{4-\tilde\varepsilon}$ for
$\tilde\varepsilon\neq4$ (with natural domain
$\text{Cyl}_{\tilde\varepsilon}^{+}\cup
\text{Cyl}_{4-\tilde\varepsilon}^{-}$), turns out to
be a symmetric trace class operator. We obtain the
same conclusion in the particular case
$\tilde\varepsilon=4$ where, starting with the
operator $\alpha\widehat\Omega^2$ defined in $\mathcal
H^+_{4}\oplus\mathcal H^-_{-4}$, we have to define its
action on $|v=0\rangle$, e.g. by annihilation, since
the operator $H'_\text{APS}$ does not decouple the
state $|v=0\rangle$. Taking into account that
$H'_\text{APS}$ was already proven to be essentially
self-adjoint \cite{kale}, a well known theorem by Kato
and Rellich \cite{kato} ensures that so is
$\widehat\Omega^2$ as well.

In order to show that its restriction to, e.g.,
$\mathcal H^+_{\tilde\varepsilon}$ is also essentially
self-adjoint, we will apply that, if $\widehat A$ is a
symmetric operator defined in certain Hilbert space
$\mathcal H$ and $\rho$ is any non-real number, then
the operator is essentially self-adjoint if and only
if there exists no solution $|\phi\rangle\in\mathcal H$
to the so-called \emph{deficiency index equation},
$\widehat A^\dagger|\phi\rangle=\rho|\phi\rangle$
\cite{GP}. Let us suppose that $\widehat\Omega^2$
defined in $\mathcal H^+_{\tilde\varepsilon}$ were not
essentially self-adjoint; this would mean that there
exists a non-trivial solution to its deficiency index
equation belonging to $\mathcal
H^+_{\tilde\varepsilon}$, which in turn would provide
a normalizable solution (identically vanishing in
$\mathcal H^-_{4-\tilde\varepsilon}$) when the
operator is defined in the larger Hilbert space
$\mathcal H^+_{\tilde\varepsilon}\oplus\mathcal
H^-_{4-\tilde\varepsilon}$. We would then reach a
contradiction because we already know that the
operator is essentially self-adjoint in this larger
space. Therefore, $\widehat\Omega^2$ has to be
essentially self-adjoint in $\mathcal
H^+_{\tilde\varepsilon}$, as we wanted to prove.

On the other hand, it was shown in Ref.
\cite{kale} that the essential and the absolutely
continuous spectra \cite{GP} of the operator
$H'_\text{APS}$ are both $[0,\infty)$. Once again,
Kato's perturbation theory \cite{kato} allows us to
extend these results to our operator
$\widehat\Omega^2$ defined in $\mathcal
H^+_{\tilde\varepsilon}\oplus\mathcal
H^-_{4-\tilde\varepsilon}$, since (up to a global
factor) it differs from $H'_\text{APS}$ in a symmetric
trace class operator. In addition, taking into account
the symmetry of $\widehat\Omega^2$ under a flip of
sign in $v$ [$f_\pm(-v)=-f_\mp(v)$] and assuming the
independence of the spectrum in the label
$\tilde\varepsilon$, we conclude that the operator
$\widehat\Omega^2$ defined in $\mathcal
H^+_{\tilde\varepsilon}$ is a positive (essentially)
self-adjoint operator whose essential and absolutely
continuous spectra are $[0,\infty)$ as well. Besides,
as we will see in Subsec. \ref{subsec:wdw}, the
(generalized) eigenfunctions of $\widehat\Omega^2$
converge for large $v$ to eigenfunctions of the
WDW counterpart of the operator. This fact, together
with the continuity of the spectrum in
geometrodynamics, suffices to conclude that the
discrete and singular spectra are empty \cite{WK}.

\subsection{Generalized Eigenfunctions}
\label{subsec:eigenstates}

Let $|e^{\tilde\varepsilon}_{\lambda}\rangle=
\sum_{v\in\mathcal L^+_{\tilde\varepsilon}}
e^{\tilde\varepsilon}_{\lambda}(v)|v\rangle$ denote a
generalized eigenstate of $\widehat\Omega^2$
corresponding to the generalized eigenvalue
$\lambda\in[0,\infty)$. For all $n\in \mathbb{N}^+$,
each coefficient
$e^{\tilde\varepsilon}_{\lambda}(\tilde\varepsilon+4n)$
of this generalized eigenfunction turns out to be
determined by the \emph{single} initial datum
$e^{\tilde\varepsilon}_{\lambda}(\tilde\varepsilon)$
in the following manner
\begin{align}\label{eigenstates}
e^{\tilde\varepsilon}_{\lambda}(\tilde\varepsilon+4n)&=
\left[\mathcal S_{\tilde\varepsilon}(0,2n)
+\frac{F(\tilde\varepsilon)}{G_\lambda(\tilde
\varepsilon-2)}\mathcal
S_{\tilde\varepsilon}(1,2n)\right]e^{\tilde\varepsilon}_{\lambda}(\tilde\varepsilon),
\end{align}
where
\begin{align}\label{FG}
F(v)&:=\frac{f_-(v)}{f_+(v)}, \qquad G_\lambda(v):=
-\frac{i\sqrt{\lambda}}{f_+(v)},
\end{align}
and
\begin{align}\label{S}
\mathcal
S_{\tilde\varepsilon}(a,b)&:=\sum_{O(a\rightarrow b)}
\left[\prod_{\{r_p\}}
F(\tilde\varepsilon+2r_p+2)\prod_{\{s_q\}}
G_\lambda(\tilde\varepsilon+2s_q) \right].
\end{align}
Here, $O(a\rightarrow b)$ denotes the set of all
possible ways to move from $a$ to $b$ by jumps of one
or two unit steps. For each element in $O(a\rightarrow
b)$, $\{r_p\}$ is the subset of integers followed by a
jump of two units, whereas $\{s_q\}$ is the subset of
integers followed by a jump of only one unit. Note
that $F(\tilde\varepsilon)=0$ for all
$\tilde\varepsilon\leq2$, so that in these cases the
second term in Eq. \eqref{eigenstates} does not
contribute.

As we stated above, the spectrum of $\widehat\Omega^2$
is positive and absolutely continuous. In terms of a
basis of generalized eigenstates
$|e^{\tilde\varepsilon}_{\lambda}\rangle$, the
spectral resolution of the identity $\mathbb{I}$ in
the kinematical Hilbert space $\mathcal
H^+_{\tilde\varepsilon}$ is given by
\begin{equation}\label{identity}
\mathbb{I}=\int_{\mathbb{R^+}} d\lambda
|e_{\lambda}^{\tilde\varepsilon}\rangle \langle
e_{\lambda}^{\tilde\varepsilon}|.
\end{equation}
Note that the integral runs just over the positive
semiaxis and the spectrum is non-degenerate. The
eigenfunctions satisfy the $\delta$-normalization
condition $\langle e^{\tilde\varepsilon}_{\lambda}
|e^{\tilde\varepsilon}_{\lambda'}\rangle=
\delta(\lambda-\lambda')$. This condition fixes the
norm of
$e^{\tilde\varepsilon}_{\lambda}(\tilde\varepsilon)$
in Eq. \eqref{eigenstates}. The only remaining freedom
is then the phase of this initial datum. We finally
fix this phase by taking
$e^{\tilde\varepsilon}_{\lambda}(\tilde\varepsilon)$
positive. The generalized eigenfunctions that form the
basis are then real, a consequence of the fact that
the difference operator $\widehat{\Omega}^2$ has real
coefficients.

It is worth emphasizing that we have been able to
solve the general eigenvalue equation of our
gravitational constraint operator $\widehat\Omega^2$,
determining explicitly the form of its generalized
eigenfunctions. This contrasts with the level of
resolution achieved in Ref. \cite{aps3}, where the
generalized eigenfunctions of the corresponding
gravitational constraint operator were given in an
iterative form and generated numerically. In this
respect, we are in an optimal situation to progress in
the comprehension of our system since we can now study
its behavior analytically. In comparison with the
exactly solvable model of Ref. \cite{acs}, here we do
not need to introduce simplifications in the system
nor restrict the study to a particular sector of
superselection, but our results are completely
general. Remember that the construction of Ref.
\cite{acs} was applied to a simplified version of the
model and only in a specific sector whose support is
centered symmetrically around $v=0$.

\subsection{Wheeler-DeWitt Limit}
\label{subsec:wdw}

Another important issue that we want to investigate is
the behavior of the quantum physical states (which we
will determine in Sec. \ref{sec:phys}) in the region
of large volume. In particular, we want to discuss
whether one recovers in that region the standard
quantization performed in geometrodynamics, namely the
WDW theory, whose predictions (for expectation values)
in turn agree on semiclassical states with the
classical ones obtained from General Relativity. To
carry out such analysis, we will only need to know how
the eigenfunctions of the gravitational constraint
operator $\widehat\Omega^2$ behave in the large $v$
limit, since the other operator involved in the
densitized Hamiltonian constraint, $\hat{p}_\phi^2$,
has already been quantized in terms of the standard
``Schroedinger-like'' representation. In this
subsection, we will obtain the eigenfunctions of the
WDW analog of the operator $\widehat\Omega^2$ and
relate them with the eigenfunctions
$e^{\tilde\varepsilon}_{\lambda}(v)$ for large $v$.

As above for LQC, in the WDW quantization we work in
the triad representation. The gravitational part of
the kinematical Hilbert space of the WDW quantization
can then be chosen as the space of square integrable
functions of $v$ with respect to the Lebesgue measure.
The operator $\hat{p}$ acts by multiplication by the
factor $p=\text{sign}(v) (2\pi\gamma
l_{\text{Pl}}^2\sqrt{\Delta}|v|)^{2/3}$, just as in
the loop quantization, and the connection is
represented by the derivative operator $\hat
c=i\,2(2\pi\gamma
l_{\text{Pl}}^2/\Delta)^{1/3}|v|^{1/6}{\partial}_{v}
|v|^{1/6}$, so that $[\hat
c,\hat{p}]=i\hbar\widehat{\{c,p\}}$.

Let us denote by $\widehat{\underline\Omega}^2$ the
operator counterpart of the classical quantity
$(cp)^2$ in the WDW theory [defined in the
Schwartz space $\mathcal{S}(\mathbb{R})$]. Since we
want to compare its features with those of
$\widehat\Omega^2$, we choose for it the analog factor
ordering, which gives rise to a symmetric operator
that is well defined in the distributional sense:
\begin{align}
\widehat{\underline\Omega}^2&=-\beta^2\sqrt{|v|}
\left[\text{sign}(v)\partial_{v}+\partial_
{v}\text{sign}(v)\right]|v|\left[\text{sign}(v)
\partial_{v}+\partial_{v}\text{sign}(v)\right]
\sqrt{|v|}=-\beta^2\left[1+4v\partial_{v}+
4(v\partial_v)^2\right],
\end{align}
where $\beta:=4\pi\gamma l_{\text{Pl}}^2$. Note
that we have simplified the expression of this
operator by disregarding the non-contributing term
$|v|\delta(v)$ in the second equality.

Owing to well known properties of the operator
$-iv\partial_v$, we can ensure that
$\widehat{\underline\Omega}^2$ is not only essentially
self-adjoint in $L^2(\mathbb{R},dv)$, but also in each
of the subspaces $L^2(\mathbb{R}^\pm,dv)$. Hence, its
action on the positive semiaxis $v>0$ and the negative
one $v<0$ are decoupled, similarly to what happens
with its analog $\widehat{\Omega}^2$ in LQC, and we
can restrict the study to $L^2(\mathbb{R}^+,dv)$.
Furthermore, in this latter Hilbert space, the
spectrum of $\widehat{\underline\Omega}^2$ is positive
and absolutely continuous. Its generalized
eigenfunctions, corresponding to any generalized
eigenvalue $\lambda\in[0,\infty)$, can be labeled by
$\sigma:=\pm\sqrt{\lambda}\in\mathbb{R}$ and are given
by
\begin{equation}\label{eq:WDW-eig}
\underline{e}_{\sigma}(v)= \frac{1}{\sqrt{2\pi\beta
|v|}}\exp\left({-i\sigma\frac{\ln{|v|}}{\beta}}\right).
\end{equation}
They provide a basis for $L^2(\mathbb{R}^+,dv)$,
normalized so that $\langle \underline{e}_{\sigma} |
\underline{e}_{\sigma^{\prime}}\rangle
=\delta(\sigma-\sigma^{\prime})$ (with $\delta$ being
the Dirac delta on the real line).

We see that the spectrum of
$\widehat{\underline\Omega}^2$ has two-fold
degeneracy, while we have shown that the spectrum of
$\widehat{\Omega}^2$ is non-degenerate. Therefore, any
loop eigenfunction converges in the large $v$ limit to
a linear combination of the two corresponding WDW
eigenfunctions (actually, one can rigorously prove
that the limit of the loop eigenfunctions indeed
exists \cite{mmp2}). Moreover, since the loop
eigenfunctions are real, both WDW components must
contribute with equal amplitude in that linear
combination. Namely, the WDW limit has the form
\begin{equation}\label{wdw-limit}
e^{\tilde{\varepsilon}}_{\lambda}(v) \to r \left\{
\exp{[i\phi_{\tilde\varepsilon}(\sigma)]}
\,\underline{e}_{\sigma}(v) +
\exp{[-i\phi_{\tilde\varepsilon}(\sigma)]}
\,\underline{e}_{-\sigma}(v) \right\},
\end{equation}
where $r$ is certain real number. In turn, one can
check numerically that the phase shift
$\phi_{\tilde\varepsilon}(\sigma)$ has the following
behavior
\begin{equation}\label{alpha-form}
\phi_{\tilde\varepsilon}(\sigma) = T(|\sigma|) +
c_{\tilde{\varepsilon}} +
R_{\tilde{\varepsilon}}(|\sigma|).
\end{equation}
where $T$ is a function of $|\sigma|$ only,
$c_{\tilde{\varepsilon}}$ is a constant which
depends on $\tilde{\varepsilon}$, and
$\lim_{\sigma\to\infty}
R_{\tilde{\varepsilon}}(|\sigma|) = 0$ \cite{mmp2}.
So, whereas the eigenfunctions of our operator are
determined by a single piece of initial data, they
behave as eigenfunctions of a second order
differential operator in the large $v$ limit,
therefore picking up a particular linear combination
of the solutions to the eigenvalue problem of that
differential operator. This nice feature of the
polymeric quantization is the main responsible of the
\emph{quantum bounce} picture, together with the
no-boundary description realized with our
superselection sectors, as we will discuss in Sec.
\ref{sec:phys2}. We postpone also to that section the
comparison between the WDW limit of our theory and
that of previous quantizations of the model.

\subsection{Operator $\widehat\Omega$}
\label{subsec:root}

Once we have characterized the gravitational
constraint operator $\widehat\Omega^2$, let us relate
it with $\widehat\Omega$ to point out some interesting
features which are due to the polymeric quantization
and that do not appear in the analog WDW theory. We
recall that (up to a multiplicative constant
factor arising from a change in the basic
commutators) $\widehat\Omega$ coincides in fact with
the operator $\widehat\Theta_i$ extensively studied in
Ref. \cite{mmp}. We now summarize its properties. Like
$\widehat\Omega^2$, it is a difference operator, but
with a constant step of two units in $v$ instead of
four. Its action is
\begin{equation}\label{actomega}
\widehat\Omega|v\rangle=-i
\big[f_+(v)|v+2\rangle-f_-(v)|v-2\rangle\big],
\end{equation}
with $f_\pm(v)$ defined in Eq. \eqref{f}. Taking into
account that $f_-(v)=0$ if $v\in(0,2]$ and $f_+(v)=0$
when $v\in[-2,0)$, we see that this operator does not
mix states $|v\rangle$ with $v$ belonging to
different semilattices of the form ${}^{(2)}\mathcal
L_{\varepsilon}^\pm:=\{v=\pm(\varepsilon+2n),\,n
\in\mathbb{N}\}$, where $\varepsilon\in(0,2]$. In
particular, our operator is well defined (with a
natural choice of domain) in the Hilbert space
${}^{(2)}\mathcal H^+_{\varepsilon}:=\mathcal
H^+_{\tilde\varepsilon=\varepsilon}\oplus\mathcal
H^+_{\tilde\varepsilon=\varepsilon+2}$. Furthermore,
from the properties of $\widehat\Omega^2$, we infer
that $\widehat\Omega$ defined in ${}^{(2)}\mathcal
H^+_{\varepsilon}$ is an essentially self-adjoint
operator whose spectrum is absolutely continuous,
non-degenerate, and equal to the real line. In turn,
its generalized eigenstates
$|e_\sigma^\varepsilon\rangle$, with support in
${}^{(2)}\mathcal L_{\tilde\varepsilon}^+$ and
corresponding to the generalized eigenvalue
$\sigma:=\pm\sqrt{\lambda}\in\mathbb{R}$, are formed
by the direct sum of two generalized eigenstates of
the squared operator $\widehat\Omega^2$ for the
eigenvalue $\lambda$, one with support in the
semilattice of step four $\mathcal
L_{\tilde\varepsilon=\varepsilon}^+$ and the other
supported in the semilattice $\mathcal
L_{\tilde\varepsilon=\varepsilon+2}^+$. Explicitly one
can see that
$|e_{\sigma}^{\varepsilon}\rangle=\sqrt{|\sigma|}[
|e_{\lambda}^{\varepsilon}\rangle\oplus
i\,\text{sign}(-\sigma)|e_{\lambda}^{\varepsilon+2}
\rangle]$ for $\sigma\neq0$, with $\langle
e_{\sigma}^{\varepsilon}|e_{\sigma'}^{\varepsilon}\rangle=
\delta(\sigma-\sigma')$ (like in the WDW case). For
$\sigma=0$, we define
$|e_{\sigma=0}^{\varepsilon}\rangle=
|e_{\lambda=0}^{\varepsilon}\rangle$ \cite{mmp}.

Let us comment that, whereas the eigenfunctions of
$\widehat\Omega^2$ have a well defined continuum limit
for large $v$, those of $\widehat\Omega$ do not
possess such a limit, since they are formed by two
components, each of them admitting a WDW limit, but
which are shifted by a phase equal to $\pm\pi/2$
(owing to the factor $\pm i$ in the linear
combination). As a consequence, when $v$ varies in
${}^{(2)}\mathcal L_{\tilde\varepsilon}^+$, the
eigenfunctions $e_{\sigma}^{\varepsilon}(v)$ oscillate
rapidly. This behavior is not present in the standard
WDW theory, where the eigenfunctions of
$\widehat{\underline\Omega}$ are continuous in $v$ for
$v>0$ [they coincide with those of
$\widehat{\underline\Omega}^2$ given in Eq.
\eqref{eq:WDW-eig}].

\section{Physical Hilbert Space} \label{sec:phys}

We can now complete the quantization by obtaining the
physical Hilbert space and providing a complete set of
observables.

The matter term present in the densitized Hamiltonian
constraint has been treated in a standard
non-polymeric way. The (essentially) self-adjoint
operator $\hat{p}_\phi^2$ is positive with a two-fold
degenerate spectrum. Its generalized eigenvalues are
labeled by $\omega^2$, with $\omega\in\mathbb{R}$. Let
us call $\mathcal{U}$ the dense domain of definition
of $\widehat{\mathcal{C}}$, invariant under its
action, from which one obtains the self-adjoint
extension of this constraint operator ($\mathcal{U}$
is the tensor product of, e.g.,
$\text{Cyl}_{\tilde\varepsilon}^{+}$ and a suitable
domain for $\hat{p}_\phi^2$). Starting from this
invariant domain, we can apply the group averaging
method to find the physical Hilbert space
$\mathcal{H}^{\textrm{Phy}}_{\tilde\varepsilon}$
\cite{gave,gave2}. The resulting physical states have
the form
\begin{align}
 \Psi(v,\phi)&=\int_0^\infty
\frac{d\lambda}{\omega(\lambda)}
e^{\tilde\varepsilon}_{\lambda}(v)\left\{\psi_+(\lambda)
\exp{\left[i\omega(\lambda)\phi\right]}+\psi_-(\lambda)
\exp{\left[-i\omega(\lambda)\phi\right]}\right\},
\end{align}
where $\psi_+(\lambda)$ and $\psi_-(\lambda)$ belong
to the physical Hilbert space
$\mathcal{H}^{\textrm{Phy}}_{\tilde\varepsilon}
=L^2(\mathbb{R}^+,\omega^{-1}(\lambda)d\lambda)$ and
$\omega(\lambda)=\sqrt{\alpha\lambda}$ [with
$\alpha=3/(4\pi\gamma^2 l_\text{Pl}^2\hbar)$].
Regarding $\phi$ as the internal time, we see that the
solutions can be decomposed in positive ($+$) and
negative ($-$) frequency components,
$\Psi_\pm(v,\phi)$, which are determined by
the initial data $\Psi_\pm(v,\phi_0)$ through the unitary evolution
$\Psi_\pm(v,\phi)=U_\pm(\phi-\phi_0)\Psi_\pm(v,\phi_0)$,
where \be U_\pm(\phi-\phi_0)=\exp{\left[\pm
i\sqrt{\alpha\,\widehat{\Omega}^2}(\phi-\phi_0)
\right]}. \ee

A complete set of observables that allows us to
interpret the system in an evolution picture is given
by the constant of motion $\hat{p}_{\phi}$ and the
relational observable $\left.\hat{v}\right|_{\phi_0}$
(or $\widehat{\left|v\right|}_{\phi_0}$ if we do not
restrict to $v>0$, see Ref. \cite{aps3}). The latter
measures the value of the volume when the time takes
the value $\phi_0$. These Dirac observables preserve
the positive and negative frequency sectors, so that,
apart from the superselection already discussed, there
exists further superselection with respect to the
frequency. We can hence restrict the study, for
instance, to the positive frequency sector.

\section{Discussion and Comparison
with Previous Work}\label{sec:phys2}

In this section, we discuss the consequences of the
quantization presented here as well as the main
similarities and differences with respect to previous
works on isotropic LQC, especially Ref. \cite{aps3},
where the physical Hilbert space was originally
determined and the quantum bounce of states that are
semiclassical at late times was first studied
satisfactorily.

Let us remember that the analog of the operator
$\widehat\Omega^2$ in Ref. \cite{aps3}, denoted by
$\Theta$, is also (essentially) self-adjoint and
positive. Furthermore, the quantum model constructed
with $\Theta$ has superselection sectors supported in
the lattices of step four $\mathcal
L_{\pm|\epsilon|}:=\{v=\pm|\epsilon|+4n,\,n\in\mathbb{Z}\}$,
where $|\epsilon|\in [0,2]$. As we see, these lattices
extend over the real line. In these superselection
sectors, the spectrum of $\Theta$ is absolutely
continuous and two-fold degenerate. Associated with
this operator, a particular ($\delta$-)orthonormal
basis of generalized eigenfunctions was chosen in Ref.
\cite{aps3}. Their elements were denoted by $e_{k}(v)$
with $k\in\mathbb{R}$, so that $e_{|k|}(v)$ and
$e_{-|k|}(v)$ have the same eigenvalue. This basis is
defined in such a way that $e_{-|k|}(v)$ tends to
$\underline{e}_{-|k|}(v)$ for large positive $v$,
where $\underline{e}_{-|k|}(v)$ is the generalized
eigenfunction of the WDW operator analog of $\Theta$
that provides an expanding wave. Similarly, we call
$\underline{e}_{|k|}(v)$ the WDW eigenfunction
corresponding to a contracting wave. The asymptotic
behavior of $e_{-|k|}(v)$ turns out to be given then
by
\begin{align}\label{wdw-limit-aps}
e_{-|k|}(v)
\xrightarrow{v\gg 1}\underline e_{-|k|}(v),\qquad
e_{-|k|}(v) \xrightarrow{v\ll -1}A
\underline{e}_{-|k|}(v)+B \underline{e}_{|k|}(v).
\end{align}
Numerical analysis has shown that, for large $|k|$,
$A$ and $B$ satisfy $|A|^2-|B|^2=1$
and $|A|\sim |B|\gg1$. As a result, the eigenfunctions
$e_{-|k|}(v)$ suffer an amplification in the negative
semiaxis. In addition, this amplification is stronger
as $|k|$ increases.

On the other hand, since the parity transformation
$v\to -v$ is a (large) gauge symmetry of the theory,
in Ref. \cite{aps3} the analysis was reduced to the
symmetric sector. Then, in general, it was necessary
to join two different lattices so that the support of
the states was symmetrically distributed around $v=0$.
Thanks to the introduction of that symmetry, the
analysis could be restricted to the positive
$v$-semiaxis.

Besides, in that work, the study was limited to the
most interesting physical states: those which are
semiclassical at late times. These states are provided
by Gaussian profiles peaked at a large momentum of the
scalar field or, equivalently, at a large negative
$k$. Thus, in such regime the contribution of
$e_{|k|}(v)$ to the physical solutions is negligible.
In this situation the form of the generalized
eigenfunctions $e_{|k|}(v)$ is irrelevant, and it was
not necessary to calculate them.

Once the parity symmetry is introduced and physical
states are restricted in practice to the region of
large negative $k$'s, one attains in Ref. \cite{aps3}
a similar scenario to ours, in the sense that, for
large positive $v$, the symmetric eigenfunctions that
contribute significantly have a WDW limit in
\emph{each} lattice $\mathcal L_{\pm|\epsilon|}$ which
is approximately of standing-wave type. This is just a
consequence of Eq. \eqref{wdw-limit-aps} and the
commented properties of the coefficients $A$ and $B$,
together with the implementation of the parity
transformation. However, it is worth emphasizing that,
while this standing-wave behavior is just an
approximation valid for $k\ll -1$ in the case of Ref.
\cite{aps3}, in our model this behavior is reached
\emph{exactly} and \emph{for all} the eigenvalues of
the scalar field momentum in the WDW limit.

As for the procedure of Ref. \cite{aps3} to restrict
to states in the parity symmetric sector, which leads
to the mentioned union of two different lattices for
generic $|\epsilon|$ (namely, $|\epsilon|\neq 0, 2$),
this has some consequences which deserve special
comment. Even when restricting the analysis to the
sector $k\ll -1$, the WDW limit of the eigenfunctions
carries a constant phase shift which depends on
$\epsilon$. Actually, one can see numerically that the
relative phase of the coefficients $A$ and $B$
presents the same kind of dependence found for
our model in Eq. \eqref{alpha-form} (with
$\tilde\varepsilon$ and $|\sigma|$ now replaced with
$\epsilon$ and $|k|$). Therefore, even in the region
of interest $k\ll -1$, two different lattices possess
different WDW limits and then their union does not
admit a global limit. Remarkably, for the
semiclassical states considered in Ref. \cite{aps3},
which are peaked for $v\gg 1$ around two classical
trajectories that do not overlap (one of an expanding
universe and the other of a contracting one), the
difference in the WDW limit of the two lattices is
just a global phase for each of the two mentioned
branches, which neither affects the norm of the state
nor the expectation values of the observables.
Nonetheless, this property of the WDW limit is not
valid for more general states.

The model studied in Ref. \cite{aps3} was later
simplified in Ref. \cite{acs}, mainly by disregarding
the quantum corrections associated with the inverse
volume operator. This simplification permitted to
obtain an exactly solvable model in which the quantum
bounce was shown in fact to be generic, although this
result was attained, however, only for a specific
superselection sector, namely, the one containing the
state $|v=0\rangle$.

In comparison with Refs. \cite{aps3,acs}, a
distinction of our proposal is that all the
superselection sectors have a support contained in a
single semiaxis. This allows us to restrict the study
to e.g. $v\in\mathbb{R}^+$ in a natural way, without
the need to appeal to a symmetrization process such as
the one described above. In any case, note that the
sectors of our model are perfectly compatible with the
imposition of parity symmetry, which can be directly
implemented by taking the direct sum of two sectors
with support in the union of two semilattices,
$\mathcal{L}^+_{\tilde\varepsilon}
\cup\mathcal{L}^-_{\tilde\varepsilon}$. Owing to the
simplicity of our sectors, the spectrum of
$\widehat{\Omega}^2$ is non-degenerate, what
facilitates the exact and explicit calculation of the
whole basis of generalized eigenfunctions [see Eq.
\eqref{eigenstates}]. An analytical and numerical
advantage of this non-degeneracy is that, to fix each
eigenfunction, we only need to impose the positiveness
of the initial datum $e^{\tilde\varepsilon}_{\lambda}
(\tilde\varepsilon)$, because its norm is completely
determined by the normalization condition. As we have
explained above, the WDW limit of these eigenfunctions
takes the form \eqref{wdw-limit}. It is a combination
of two WDW eigenfunctions which can be interpreted as
contracting and expanding components (in $v$), or
equivalently as incoming and outgoing components. They
contribute with equal amplitude since the
eigenfunctions are real, and in this sense the limit
is \emph{exactly} a standing wave. On the other hand,
our eigenfunctions have support in a semiaxis which
does not contain the potential singularity. This
behavior does not arise from the imposition of any
particular condition, like e.g. a boundary condition,
but it is a natural feature of our model, explainable
only by the functional properties of our gravitational
constraint operator. From this perspective, we
consider that our model provides an intrinsic
no-boundary description. This implies that the
outgoing component must evolve to an incoming
component and vice versa, since the flux cannot scape
across $v=0$. Therefore, the expanding and contracting
components must represent the two branches of a
bouncing universe. Unlike in Ref. \cite{aps3}, where
only a certain regime is considered, this result is
independent of the spectral profile of interest.
Furthermore, our analysis is valid for all choices of
superselection sector (i.e., for all values of
$\tilde{\varepsilon}$), in contrast with the
discussion carried out in Ref. \cite{acs}. In short,
we obtain a completely \textit{generic} quantum
bounce. Obviously, the commented expanding and
contracting components will be peaked (for large $v$)
around well differentiated trajectories only for
certain types of states.

Let us conclude by remarking that the kind of
no-boundary description that we have reached, which
plays a key role in the arguments leading to the
picture of a quantum bounce, is a characteristic of
our model that is not shared by any of the previous
works on flat isotropic LQC
\cite{abl,aps1,aps3,acs,kale,Pol}.

\section{Conclusion}
\label{sec:dis}

We have presented an alternative quantization in LQC
of a flat, homogeneous, and isotropic universe in the
presence of a massless scalar (homogeneous) field with
the aim of improving our understanding of the theory.
More explicitly, by a new choice of factor ordering,
motivated by studies of anisotropic cosmologies, we
have symmetrized the Hamiltonian constraint in a way
which results specially appropriate to investigate
some issues which had not been taken into account with
sufficient care in previous analyses, such us the
precise relation between the Hamiltonian constraint
and its densitized version, the possibility to attain
superselection sectors which, being as simple as
possible, posses optimal properties, and a clear
comprehension of the WDW limit of the theory.

Our Hamiltonian constraint operator presents two nice
features: under its action, the zero-volume state
decouples, and states with different orientation of
the triad are not mixed. Owing to the former of these
properties, in our model the big bang singularity is
kinematically resolved inasmuch as its quantum analog,
namely the zero-volume state, is removed from the
kinematical Hilbert space. This fact allows us to
establish a bijection between the non-trivial
solutions of the Hamiltonian constraint (with
densitization equal to that used in LQG) and those of
its densitized version (with respect to the volume).
This bijection is not established in the kinematical
Hilbert space, but in a larger space that provides a
natural home for the solutions. For the densitized
version of the constraint, the identification of some
Dirac observables is straightforward. In summary, our
analysis shows in a rigorous way how one can start
with the density weight for the Hamiltonian constraint
which arises naturally in LQG and make the passage to
a simpler constraint with different densitization.

The second feature commented above, namely the
decoupling between triads with different orientation,
has also important consequences. Thanks to it, our
Hamiltonian constraint superselects sectors with
support in a single semiaxis of the real line, instead
of the whole real line. The simplicity of our sectors
in turn simplifies considerably the construction of
the physical Hilbert space in comparison with previous
works \cite{abl,aps1,aps3,acs,kale,Pol}. In those
references the superselection sectors have support in
lattices of the real line. The study can be restricted
to the positive semiaxis $v>0$ by demanding parity
symmetry, but in the generic case this is made at the
cost of replacing the original individual lattices by
the union of two lattices that are transformed one
into each other under parity. In contrast, in our case
the functional properties of the gravitational
constraint operator restrict the study to $v>0$
directly. As a consequence, the generalized
eigenfunctions of this constraint operator, which are
the elements that account for the quantum information
about the geometry, provide a no-boundary description,
arising in a single section of minimum non-zero volume
$v=\varepsilon$ without the need to introduce any
specific (boundary or large gauge) condition in order
to affect the behavior in the vicinity of the origin.
Another related consequence is that the spectrum of
the gravitational constraint operator is
non-degenerate, whereas the spectrum of the analog WDW
operator is two-fold degenerate. Taking into account
that the eigenfunctions of the loop operator can be
chosen to be real, they must converge in the WDW limit
(large $v$ limit) to a standing wave composed of two
WDW eigenfunctions equally contributing, one outgoing
and the other incoming. In turn, the no-boundary
behavior implies that neither the incoming flux nor
the outgoing one can ``scape'' to the region $v<0$.
Thus, the only possibility is a bounce in which the
incoming component becomes outgoing and vice versa.
This shows the occurrence of a quantum bounce in a
generic manner. Our conclusions do not only confirm
the results obtained in previous works
\cite{aps3,acs}, in particular concerning the
robustness of the bounce, but also reinforce them
inasmuch as the discussed scenario is completely
general. We have neither focused the study on a
concrete class of physical states, unlike in Ref.
\cite{aps3}, nor simplified and particularized it to a
specific superselection sector, unlike in Ref.
\cite{acs}. Finally, we note that the commented bounce
scenario does not imply that the outgoing and ingoing
components peak at non-overlapping classical
trajectories, so that such a semiclassical behavior
will be reached only for certain physical states (e.g.
those considered in Ref. \cite{aps3}).

In conclusion, the simplicity of our quantum model has
allowed us to solve it explicitly, gaining the
physical intuition necessary to develop our
understanding of LQC without resorting to numerical
analyses. In this way, we have improved the control
over the WDW limit and analyzed in depth the
fundamental reasons behind the occurrence of a quantum
bounce.

\section*{Acknowledgments}

This work was supported by the Spa\-nish MICINN
Project FIS2008-06078-C03-03 and the
Consolider-Inge\-nio 2010 Program CPAN
(CSD2007-00042). The authors are grateful to D.
Brizuela, L.J. Garay, T. Paw{\l}owski, R.
Rodr\'iguez-Oliveros, and J.M. Velhinho for
discussions. M.M.-B. acknowledges financial aid by the
I3P Program of CSIC and the European Social Fund under
the grant I3P-BPD2006. She thanks A.E.I. for warm
hospitality during part of the period of development
of this work. J.O. acknowledges CSIC by financial
support under the grant JAEPre\_08\_00791.

\bibliographystyle{plain}

\begin{thebibliography}{99}
\bibitem{lqg3} T. Thiemann, {\it{Modern Canonical
Quantum General Relativity}} (Cambridge University
Press, Cambridge, England, 2007).
\bibitem{lqg1} C. Rovelli, {\it{Quantum Gravity}}
(Cambridge University Press, Cambridge, England,
2004).
\bibitem{lqg2} A. Ashtekar and J. Lewandowski,
Classical Quantum Gravity {\bf 21}, R53 (2004).
\bibitem{lqc} M. Bojowald, Living Rev. Rel. {\bf 11},
4 (2008).
\bibitem{abl} A. Ashtekar, M. Bojowald, and J. Lewandowski,
Adv. Theor. Math. Phys. {\bf 7}, 233 (2003).
\bibitem{aps1} A. Ashtekar, T. Paw{\l}owski, and P. Singh,
Phys. Rev. Lett. {\bf 96}, 141301 (2006);
Phys. Rev. D {\bf 73}, 124038 (2006).
\bibitem{aps3} A. Ashtekar, T. Paw{\l}owski, and P. Singh,
Phys. Rev. D {\bf 74}, 084003 (2006).
\bibitem{acs}A. Ashtekar, A. Corichi, and P. Singh,
Phys. Rev. D {\bf 77}, 024046 (2008).
\bibitem{cs} A. Corichi and P. Singh, Phys. Rev.
Lett. {\bf 100}, 161302 (2008); Phys. Rev. D {\bf80}, 044024 (2009).
\bibitem{kale} W. Kami\'nski and J. Lewandowski,
Classical Quantum Gravity {\bf 25}, 035001 (2008).
\bibitem{Pol} W. Kami\'nski, J. Lewandowski, and T.
Paw{\l}owski. Classical Quantum Gravity {\bf 26},
035012 (2009).
\bibitem{apsv} A. Ashtekar, T. Paw{\l}owski, P. Singh, and
K. Vandersloot, Phys. Rev. D {\bf 75}, 024035 (2007).
\bibitem{vand} K. Vandersloot, Phys. Rev. D {\bf 75}, 023523
(2007).
\bibitem{skl} L. Szulc, W. Kami\'nski, and J. Lewandowski,
Classical Quantum Gravity {\bf 24}, 2621 (2007).
\bibitem{tom} E. Bentivegna and T. Paw{\l}owski,
Phys. Rev. D {\bf 77}, 124025 (2008).
\bibitem{chio} D.W. Chiou, Phys. Rev. D {\bf 75},
024029 (2007).
\bibitem{mmp} M. Mart\'{\i}n-Benito, G.A. Mena Marug\'{a}n,
and T. Paw{\l}owski, Phys. Rev. D {\bf 78}, 064008
(2008).
\bibitem{mmp2}M. Mart\'{\i}n-Benito, G.A. Mena Marug\'{a}n,
and T. Paw{\l}owski, Phys. Rev. D {\bf 80}, 084038
(2009).
\bibitem{awe} A. Ashtekar and E. Wilson-Ewing, Phys. Rev.
D {\bf 79}, 083535 (2009).
\bibitem{gow} M. Mart\'{\i}n-Benito, L.J. Garay, and G.A.
Mena Marug\'{a}n, Phys. Rev. D \textbf{78}, 064008
(2008); G.A. Mena Marug\'{a}n and M.
Mart\'{\i}n-Benito, Int. J. Mod. Phys. A {\bf24}, 2820
(2009).
\bibitem{boj} M. Bojowald, Classical Quantum Gravity {\bf 20},
2595 (2003).
\bibitem{note1} Here $\tau_i$ are elements of the su(2)
algebra, proportional to the Pauli matrices, and satisfy
$\tau_i\tau_j=\frac{1}{2}\varepsilon_{ijk}\tau^k-\frac{1}{4}
\delta_{ij}$.
\bibitem{Vel} J.M. Velhinho, Classical Quantum Gravity
\textbf{24}, 3745 (2007).
\bibitem{Castro} G.A. Mena Marug\'an, AIP Conf. Proceedings
{\bf 1130}, 89 (2009).
\bibitem{notepl} This would also eliminate possible
dependences on the choice of fiducial cell in the
effective theory at subleading orders. See e.g. Y.
Ding, Y. Ma, and J. Yang, Phys. Rev. Lett. {\bf 102},
051301 (2009).
\bibitem{kato} T. Kato, {\it Perturbation Theory for Linear
Operators} (Springer-Verlag, Berlin, 1980).
\bibitem{GP} See e.g. A. Galindo and P. Pascual,
{\it{Quantum Mechanics I}} (Springer-Verlag, Berlin, 1990);
M. Reed and B. Simon, {\it Methods
of Modern Mathematical Physics I: Functional Analysis}
(Academic Press, San Diego, 1980).
\bibitem{WK} W. Kami\'nski (unpublished).
\bibitem{gave} D. Marolf, \texttt{arXiv:gr-qc/9508015};
Classical Quantum Gravity {\bf 12}, 1199 (1995); {\bf 12},
1441 (1995); {\bf 12}, 2469, (1995).
\bibitem{gave2} A. Ashtekar, J. Lewandowski, D. Marolf,
J. Mour\~ao, and T. Thiemann, J. Math. Phys. {\bf 36}, 6456
(1995).

\end{thebibliography}

\end{document}